\documentclass[aps,twocolumn,lettersize,showpacs,longbibliography]{revtex4-1}

\usepackage{amsmath}
\usepackage{graphicx}
\usepackage{mathtools}
\usepackage{xcolor}
\usepackage{bm}

\DeclarePairedDelimiter\floor{\lfloor}{\rfloor}

\newcommand{\costp}{\cos(2\varphi)}

\newcommand{\costps}{\cos^2(2\varphi)}
\newcommand{\sintps}{\sin^2(2\varphi)}
\newcommand{\cosp}{\cos\varphi}
\newcommand{\sinp}{\sin\varphi}
\newcommand{\ahat}{\hat{a}}
\newcommand{\adaghat}{\hat{a}^{\dagger}}

\def\bra#1{\mathinner{\langle{#1}|}}
\def\branobar#1{\mathinner{\langle{#1}}}
\def\ket#1{\mathinner{|{#1}\rangle}}

\begin{document}
\title{Hall quantization and optical conductivity evolution with variable Berry phase in $\alpha$-$T_3$ model}
\author{E. Illes$^1$}
\email{illese@uoguelph.ca}
\author{J. P. Carbotte$^{2,3}$}
\author{E. J. Nicol$^1$}
\affiliation{$^1$Department of Physics, University of Guelph, Guelph, Ontario N1G 2W1, Canada and\\
Guelph-Waterloo Physics Institute, University of Guelph, Guelph, Ontario N1G 2W1, Canada\\
$^2$Department of Physics, McMaster University, Hamilton, Ontario L8S 4M1, Canada\\
$^3$The Canadian Institute for Advanced Research, Toronto, Ontario M5G 1Z8, Canada }
\date{\today}

\begin{abstract}
The $\alpha$-$T_3$ model is characterized by a variable Berry phase that changes continuously from $\pi$ to $0$.  We take advantage of this property to highlight the effects of this underlying geometrical phase on a number of physical quantities.  The Hall quantization of the two limiting cases is dramatically different - a relativistic series is associated with a Berry phase of $\pi$ and a non-relativistic series is associated with the other limit.  We study the quantization of the Hall plateaux as they continuously evolve from a relativistic to a non-relativistic regime.  Additionally, we describe two physical quantities that retain knowledge of the Berry phase, in the absence of a motion-inducing magnetic field.  The variable Berry phase of the $\alpha$-$T_3$ model allows us to explicitly describe the Berry phase dependence of the dynamical longitudinal optical conductivity and of the angular scattering probability.    
\end{abstract}
	
\maketitle

\section{Introduction}
Graphene, a single-atom-thick sheet of carbon atoms arranged on a honeycomb lattice (HCL), was first experimentally isolated in 2004~\cite{novoselov:2004}.  It is a gapless semiconductor with low-energy excitations that are well-described by a massless two-dimensional Dirac Hamiltonian, or the Dirac-Weyl equation with pseudospin $S=1/2$.  It exhibits an unconventional quantum Hall effect (QHE)~\cite{zheng:2002,gusynin:2005_1,zhang:2005,novoselov:2007}, characterized by a half-integer shift in the Hall quantization.  This phenomenon is a direct consequence of its unusual Berry phase of $\pi$ found in graphene.   

An analogous lattice, the $T_3$ or dice lattice, is described by the same Dirac-Weyl Hamiltonian, but with pseudospin $S=1$.  Here the geometry of the honeycomb lattice is augmented by an additional atom that sits at the center of each hexagon coupled to one of the two topologically inequivalent sites of the honeycomb. This lattice can be naturally formed by growing a trilayer structure of cubic lattices in the (111) direction~\cite{wang:2011} (for example SrTiO$_3$/SrIrO$_3$/SrTiO$_3$) or by confining cold atoms to an optical lattice~\cite{bercioux:2009}.  A number of recent papers\cite{dora:2011,malcolm:2014,lan:2011} have explored the properties of general pseudospin $S$ lattices, arising from the generalized Dirac-Weyl Hamiltonian, and provide insight into lattices with pseudospin $S=1/2,1$ and beyond.

The $\alpha$-$T_3$ model~\cite{raoux:2014} provides a continuous evolution between the honeycomb ($\alpha=0$) and dice ($\alpha=1$) lattice via the parameter $\alpha$, which is proportional to the strength of the coupling with the additional atom at the center of the HCL.  Notably, a recent paper by Malcolm and Nicol ~\cite{malcolm:2015} demonstrated that at a critical doping Hg$_{1-x}$Cd$_x$Te maps onto the $\alpha$-$T_3$ model in the intermediate regime (between the dice and HCL), with an $\alpha=1/\sqrt{3}$.  The orbital magnetic response of such a lattice is particularly intriguing.  At the Dirac point it goes from diamagnetic ~\cite{mcclure:1956} ($\alpha=0$) to paramagnetic ~\cite{sutherland:1986,vidal:1998} ($\alpha=1$).  This behaviour has recently been linked to the evolution of the Berry phase in this system, which changes continuously from $\pi$ to zero as it evolves from honeycomb to dice, respectively.  A more detailed discussion of the phase can be found in Ref.~\cite{louvet:2015}.

Berry's phase, discovered by Berry in 1984~\cite{berry:1984,berry:1989}, is a geometrical phase that quantum mechanical systems acquire while undergoing adiabatic transport and can impact physical quantities involving the motion of electronic charge such as optical transport phenomena and magnetic field induced orbits..  The $\alpha$-$T_3$ model is unusual in that it is an example of a model with a continuous evolution of this phase.  In this paper, we take advantage of the variable Berry phase of the $\alpha$-$T_3$ model to explore the role of this underlying geometrical phase in quantities such as the density of states, the DC Hall conductivity, the dynamical optical conductivity, and the angular scattering probability.  

The Hall conductivity is known to be dependent on the Berry phase.  Here we study how the quantization of the Hall plateaux evolves as the Berry phase of the $\alpha$-$T_3$ model varies from $\pi$ to $0$.  A Berry phase of $\pi$ is associated with a pure relativistic (Dirac) Hamiltonian with energy proportional to momentum (with Fermi velocity $v_F$) and both positive and negative energy states.  Traditionally, a Berry phase of $0$ applies to a pure non-relativistic (Schrodinger) Hamiltonian with positive states only, and energy dispersion curves quadratic in momentum and Schrodinger mass $m$.  For the case when both terms are present, Li and Carbotte ~\cite{li:2014} showed that when the Dirac term dominates the Hall quantization takes on its relativistic form.  In the other limit when the Schrodinger term dominates, the quantization is non-relativistic even when spin degeneracy is lifted by a small spin orbit coupling term.  In the work of Li and Carbotte it was not possible however to connect these two regimes in a continuous manner.  Nonetheless, it remains an interesting question to understand the evolution of the Hall quantization from a relativistic Dirac regime to a non-relativistic quantization in a continuous way.  The $\alpha$-$T_3$ model offers this possibility, though it does not contain a Schrodinger term.

The paper is structured as follows.  In section~\ref{sec:model} the $\alpha$-$T_3$ model is specified and the Hamiltonian with and without magnetic field $B$ is given.  In section~\ref{sec:density}  we calculate the density of states for the $\alpha$-$T_3$ model under a finite magnetic field, which is central to our discussion in section~\ref{sec:hall}.  The $B\rightarrow 0$ limit of the density of states is discussed and its relationship to the finite $B$ case is stressed.   In section~\ref{sec:hall} we derive the Hall quantization from the grand potential via the magnetization.  Its relationship to the underlying Berry phase is emphasized.  In section~\ref{sec:optics} we consider dynamical longitudinal conductivity $\sigma_{xx}(\omega)$ in the $B=0$ limit and show how, in sharp contrast to what we found for the DOS in this same limit, it does depend on the Berry phase, as does the backscattering amplitude in section~\ref{sec:scattering}.  Conclusions are found in section~\ref{sec:conclusion}.

\section{The $\alpha$-$T_3$ Model}\label{sec:model}
\begin{figure}[tb]
\centering
 \includegraphics[width=\columnwidth]{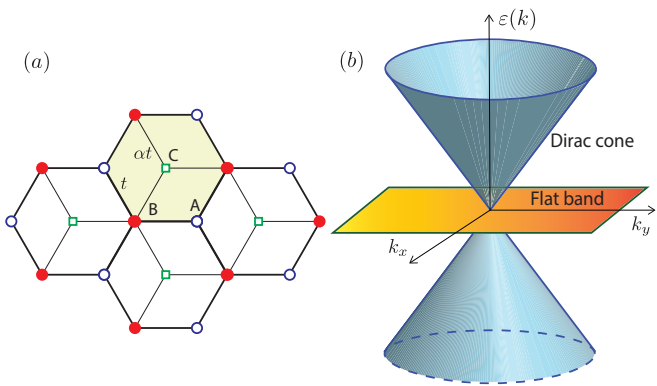}
\caption{(Color online) Panel (a) shows the lattice of the $\alpha$-$T_3$ model with hopping $t$ between the atoms in the HCL and hopping $\alpha t$ between the atoms at the $B$ and $C$ sites.  Panel (b) is a schematic diagram of the energy dispersion at a single $K$ point, with cones for the linearly dispersing valence and conduction band, and a dispersionless flat band at zero energy that cuts through the Dirac point.}
\label{fig:intro}
\end{figure}
\begin{figure}[tb]
\centering
 \includegraphics[width=\columnwidth]{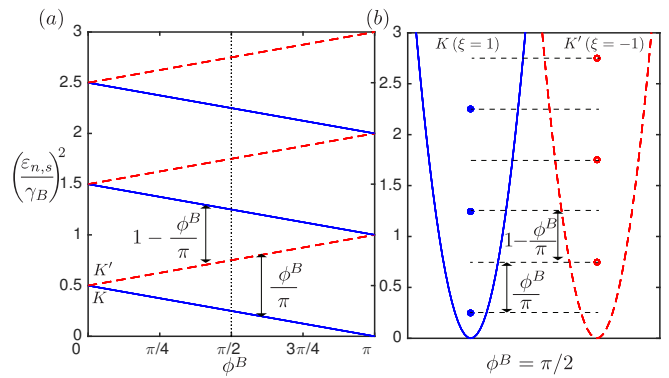}
\caption{(Color online) Panel (a) depicts the smooth evolution of the square of the Landau level energies as a function of the Berry phase for the cones as calculated from the square of Eq.~\eqref{eq:energy1}.  The $K$ and $K^{\prime}$ valleys are shown in red-dashed and solid blue respectively for $n=1,2,3$.  Panel (b) highlights the case of $\phi^B=\pi/2$.  Here, the square of the LL energies are schematically represented by dots, and the parabolas are a schematic representation of the square of the Dirac cones.  In both panels, arrows denote the Berry phase dependent separation between the square of the LLs.}
\label{fig:berry}
\end{figure} 
The $\alpha$-$T_3$ model interpolates between the pseudospin $S$=1/2 honeycomb lattice (HCL) of graphene, and the pseudospin $S$=1 dice (or $T_3$) lattice via the parameter $\alpha$.  Fig.~\ref{fig:intro}(a) depicts the $\alpha$-$T_3$ lattice, with three atoms per unit cell, where sites $A$ and $B$ make up the HCL with hopping amplitude $t$, and site $C$ sits at the center of the hexagon, connected only to site $B$ with hopping amplitude $\alpha t$. We will refer to $\alpha=0$ as graphene, and $\alpha=1$ as the dice lattice throughout this paper.  The low-energy tight-binding Hamiltonian for the $\alpha$-$T_3$ model~\cite{raoux:2014} is given by
\begin{equation}\label{eq:hamiltonian1}
H(\pmb{k}) = \left(\begin{array}{c c c} 0 & f_{\pmb{k}}\cosp & 0 \\
f^*_{\pmb{k}}\cosp & 0 & f_{\pmb{k}}\sinp \\ 
0 & f^*_{\pmb{k}}\sinp & 0 \\
\end{array}\right)
\end{equation}
with $f_{\pmb{k}}=\hbar v_F(\xi k_x-ik_y)$, and $\xi=\pm1$ a valley index for the $K$ and $K^{\prime}$ valleys, respectively, associated with the two-dimensional Brillouin zone. The angle $\varphi$ is related to the strength of the coupling $\alpha$ as $\alpha=\tan\varphi$, and the Hamiltonian has been rescaled by $\cosp$ for convenience.   

There are three atoms per unit cell in the $\alpha$-$T_3$ lattice, and therefore three bands: a flat band with energy $\varepsilon_{\pmb{k},0}=0$ for all momenta, and two cones with energy $\varepsilon_{\pmb{k},s}=s\hbar v_F k$, where $s=\pm1$, for the conduction and valence band, respectively.  The low-energy spectrum of the $\alpha$-$T_3$ lattice is depicted in Fig.~\ref{fig:intro}(b) showing a single $K$ point with two linearly dispersing bands (Dirac cones) and a dispersionless flat band with zero energy.  All three bands are present for the full range of $\alpha$.  For $\alpha=0$, the atoms at site $C$ are uncoupled from the atoms at site $B$.  The respective wavefunctions are
\begin{equation}\label{eq:wf1}
\ket{\Psi_{0}} =\left(\begin{array}{c } \sinp\, e^{i\theta_{\pmb{k}}} \\
 0 \\ 
-\cosp\, e^{-i\theta_{\pmb{k}}}\\
\end{array} \right)
\end{equation}
for the flat band, and 
\begin{equation}\label{eq:wf2}
\ket{\Psi_{s}} =\frac{1}{\sqrt{2}}\left(\begin{array}{c } \cosp\, e^{i\theta_{\pmb{k}}} \\
 s \\ 
\sinp\, e^{-i\theta_{\pmb{k}}}\\
\end{array} \right),\\
\end{equation}
for the conduction and valence band.  Here, $\theta_{\pmb{k}}$ is the angle associated with momentum $\pmb{k}$ such that $f_{\pmb{k}}=|f_{\pmb{k}}|e^{i\theta_{\pmb{k}}}$. The Berry phase of an orbit in the conical bands is 
\begin{align}
\phi_{\xi}^B=\pi\xi\costp=\pi\xi\left(\frac{1-\alpha^2}{1+\alpha^2}\right)
\end{align} and 
\begin{align}
\phi_{0,\xi}^B=-2\pi\xi\costp=-2\pi\xi\left(\frac{1-\alpha^2}{1+\alpha^2}\right)
\end{align}
for the flat band, in terms of the coupling $\alpha$.  With the exception of $\alpha=0,1$ the Berry phase is different in the $K$ and $K^{\prime}$-valleys.  It is non-topological~\cite{louvet:2015} and is a smooth function of $\alpha$, going from  graphene ($\phi_{\xi}^B=\pi$) to the dice lattice ($\phi_{\xi}^B=0$).  

When a magnetic field $B$ is applied perpendicular to the plane of the crystal lattice, the resulting Hamiltonian takes the form
\begin{equation}
H_K= -H^*_{K^{\prime}}= \gamma_B \left(\begin{array}{c c c} 0 & \cosp \ahat & 0 \\
\cosp \adaghat & 0 & \sinp \ahat \\ 
0 & \sinp \adaghat & 0 \\
\end{array}\right)
\end{equation}
with $\gamma_B$ a magnetic energy scale given by $\gamma_B=v_F\sqrt{2eB\hbar}$.  Here $\adaghat$ and $\ahat$ are the creation and annihilation operators, respectively, that obey the usual commutation relation $[\ahat,\adaghat]=1$ and act on Fock states such that $\adaghat\ket{n}=\sqrt{n+1}\ket{n+1}$ and $\ahat\ket{n}=\sqrt{n}\ket{n-1}$. The Landau level energies for the conduction and valence band are
\begin{align}\label{eq:energy1}
\varepsilon_{n,s}&= s \gamma_B \sqrt{n-\frac{1}{2}-\frac{\phi^B_{\xi}}{2\pi}}
\end{align}
with $n=1,2,3,...$ in terms of the Berry phase, $\phi_{\xi}^B$.  The flat band retains its energy $\varepsilon_{n,0}=0$.  

In Fig.~\ref{fig:berry}, we illustrate the Berry phase dependence of the LL energies of the $\alpha$-$T_3$ model.  Panel (a) depicts the square of the LL energies deriving from the conical bands, showing a linear dependence on the Berry phase.  They are plotted for the first three values of the index $n$ ($n=1,2,3$).  We note that the Berry phase in the $K$ and $K^{\prime}$-valleys is distinct, with the exception of the two endpoints $\phi^B=0,\pi$.  Here we have introduced $\phi^B$, the Berry phase of the valence and conduction band in a the $K$ valley, defined by $\phi^B_{\xi}=\xi{\phi^B}$ for convenience.  At $\phi^B=0$ the LL from the two valleys merge to the same energy.  For $\phi^B\rightarrow\pi$ only the $K$-valley LL approaches zero energy in this model.  The other LL energies match in pairs with the $n=m$ level of the $K$-valley matching up with the $n=m-1$ level of the $K^{\prime}$-valley, for $m$ an integer ($m=2,3,4...$).  For the intermediate cases, $0<\phi^B<\pi$, the Berry phase dependent separation between the square of the LL energies in the $K$ and $K^{\prime}$ valleys is identified by arrows.  The case of $\phi^B=\pi/2$ is highlighted in panel (b) of Fig.~\ref{fig:berry}, with arrows emphasizing the role of the Berry phase.  Here, the square of the LL energies are plotted in relation to the parabolas that result from squaring the conical energy dispersion of the associated $B=0$ bands.

\section{Density of States}\label{sec:density}
\begin{figure}[tb]
\centering
 \includegraphics[width=\columnwidth]{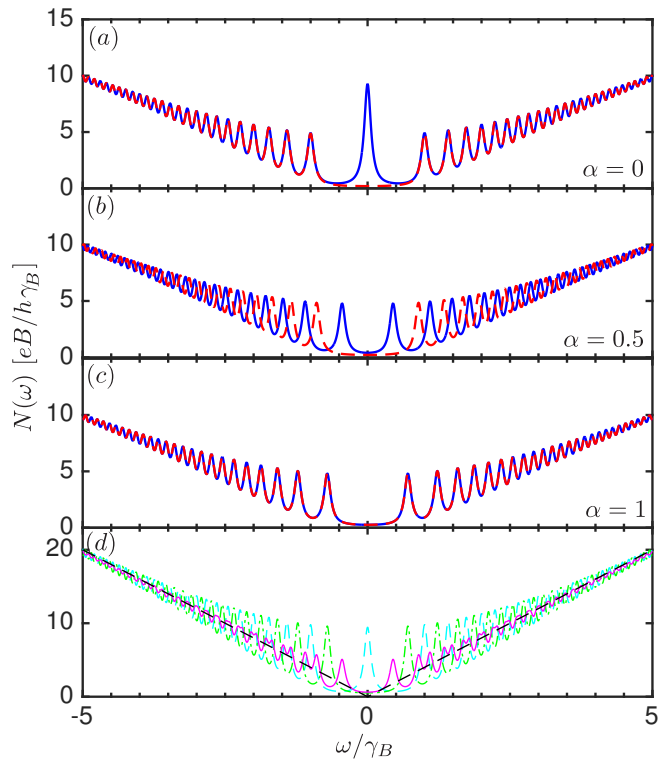}
\caption{(Color online) The density of states of the conical bands in the presence of a magnetic field showing the LLs, as calculated from Eq.~\eqref{eq:density_dim_less} with a broadening of $\Gamma=0.07\gamma_B$.  Panels (a) - (c) show density of states for the $K$-valley in solid blue, and the $K^{\prime}$-valley in red dashed.  Panel (d) shows the total density of states (summed over both valleys) for $\alpha=0,0.5,1$ in dashed light blue, solid magenta, and dot-dashed green, respectively. In our units the $B=0$ background is 2 in (a)-(c) and 4 in (d) and is shown as a black dashed line in (d).}
\label{fig:density}
\end{figure}
In this section we examine the density of states of the $\alpha$-$T_3$ model.  We draw the reader's attention to the role of Berry's phase in the density of states in the presence of a magnetic field, and contrast this to the zero field case.   

As there are three atoms per unit cell in the $\alpha$-$T_3 $ lattice we have three sets of Landau levels.  The energies of the Landau levels associated with the valence and conduction band can be found in the previous section in Eq.~\eqref{eq:energy1}.  Here, we will not be interested in the flat band, as its energy is zero for all $n$.  The contribution to the density of states $N_{\xi}(\omega)$ from the remaining two bands in the $\xi$ valley is
\begin{align}\label{eq:density1}
N_{\xi}(\omega) &= \frac{eB}{h}\sum_{n=1}^{\infty}\sum_{s=\pm}\delta\left(\omega-s\gamma_B\sqrt{n-\frac{1}{2}-\frac{\phi^B_{\xi}}{2\pi}}\right)	
\end{align}
where the prefactor $eB/h$ is the usual density of states factor in a magnetic field.  It is useful before proceeding further to compare Eq.~\eqref{eq:density1} to the density of states of graphene.  To do so, we rewrite Eq.~\eqref{eq:density1} as
\begin{align}\label{eq:density2}
N_{\xi}(\omega) =& \frac{eB}{h} \delta\left(\omega-\gamma_B\sqrt{\frac{1}{2}- \frac{\xi}{2}}\right)\nonumber\\ &+\frac{eB}{h}\delta\left(\omega+\gamma_B\sqrt{\frac{1}{2}-\frac{\xi}{2}}\right)\\
&+\frac{eB}{h}\sum_{n=2}^{\infty}\sum_{s=\pm}\delta\left(\omega-s\gamma_B\sqrt{n-\frac{1}{2}-\frac{\xi}{2}}\right)	\nonumber
\end{align}
where we have explicitly written out the terms associated with the lowest LLs and put in the appropriate Berry phase $\phi_{\xi}^B=\xi\pi$.  For the first two terms, this results in a $0$ under the square root for the $K$ valley, and a $1$ under the square root for the $K^{\prime}$ valley.   Thus, in the $\alpha$-$T_3$ model, both of the Dirac $\delta$-functions associated with zero energy LLs are contained in the $K$ valley, and none in the $K^{\prime}$ valley.  In contrast, both the $K$ and $K^{\prime}$ valleys in graphene contain one of these functions.  While nominally graphene and the $\alpha$-$T_3$ model are based on a honeycomb lattice, the additional central atom in the $\alpha$-$T_3$ model makes it a 3-atoms per unit cell situation.  Moreover, the coupling of the central atom to only the $B$ sublattice introduces the valley asymmetry seen here.  Upon summing over the two valleys, and ignoring the contribution from the flat band, the results for the DOS agree with that of graphene.

We now write the density of states in a dimensionless form
\begin{align}\label{eq:density_dim_less}
	\bar{N}_{\xi}(\bar{\omega})&=\sum_{n=1}^{\infty}\sum_{s=\pm}\delta\left(\bar{\omega}-s\sqrt{n-\frac{1}{2}-\frac{\phi_{\xi}^B}{2\pi}}\right)
\end{align}	
with $\bar{N}_{\xi}=(\gamma_Bh/eB) N_{\xi}$ and $\bar{\omega}=\omega/\gamma_B$.  The $\delta$ functions above correspond to LLs with energy $\varepsilon_{n,s}$, the separation between which is dictated by the Berry phase (see Fig.~\ref{fig:berry}).  To plot the DOS, these $\delta$-functions are broadened by scattering $\Gamma$ as 
\begin{align}
\delta(\bar{\omega}-\bar{\varepsilon}_{n,s})&=\frac{1}{\pi}\frac{\Gamma}{(\bar{\omega}-\bar{\varepsilon}_{n,s})^2+\Gamma^2},	
\end{align}
where $\bar\varepsilon_{n,s}=\varepsilon_{n,s}/\gamma_B$.  The dimensionless density of states is plotted for representative values of $\alpha$  in Fig.~\ref{fig:density}.  For graphene, in the presence of a magnetic field, LLs are observed in the density of states~\cite{guohong:2007,guohong:2009,pound:2011}.  Here, for all values of $\alpha$, we observe the expected particle hole symmetry, as well as broadened peaks centered about the LLs of the system. 

In the top panel we show the density of states for $\alpha=0$ and we observe the expected graphene behaviour, with the exception of the missing zero LL in the $K^{\prime}$ valley about $\bar{\omega}=0$.  The density of states summed over the two valleys, shown in the bottom panel of Fig.~\ref{fig:density}, does recover the expected result for graphene.  For the intermediate case of $\alpha=0.5$ the magnetic oscillations in the two valleys are shifted with respect to each other.  This is a manifestation of the offset between LLs in the two valleys as was highlighted in Fig.~\ref{fig:berry}.  In frame (c) the density of states, $\bar{N}_{\xi}(\omega)$, is identical in both valleys, as we expect, since the LL energies are the same in for $K$ and $K^{\prime}$ for $\phi^B=0$.  

It is important to note that for $\omega>>\gamma_B$ all frames in Fig.~\ref{fig:density} converge to the same background value, independent of Berry phase.  This is demonstrated in panel (d) which shows the density of states summed over the two valleys for the three representative values of $\alpha$ ($\alpha=0,0.5,1$).  In the units we used in the figure, the background is a line of slope 4.  

In the limit of small magnetic field ($B\rightarrow 0$) we recover the usual continuum density of states $N(\omega)=\omega/(2\pi \hbar^2 v_F^2)$, which is independent of the Berry phase.  This can be obtained by taking $\bar{\omega}$ very large, which results in the Dirac-delta function only contributing for large $n$.  In this limit, the constant term under the square root ($1/2 + \phi^B_{\xi}/\pi$) becomes negligible (compared to large $n$) resulting in
\begin{align}
N_{\xi}(\omega)=&\frac{eB}{h\gamma_B} 2\sqrt{N_c},	
\end{align}
with $\sqrt{N_c}=\omega/\gamma_B$.  This works out to be $N(\omega)=\omega/(2\pi \hbar^2 v_F^2)$, as we know from direct calculation in the zero field ($B=0$) case.

The density of states might suggest that Berry's phase only plays a role in the presence of a magnetic field.  However, this is not the case, as we will demonstrate with two examples in later sections.

\section{DC Hall Conductivity from Magnetization}\label{sec:hall}

\begin{figure}[tbh]
\centering
 \includegraphics[width=\columnwidth]{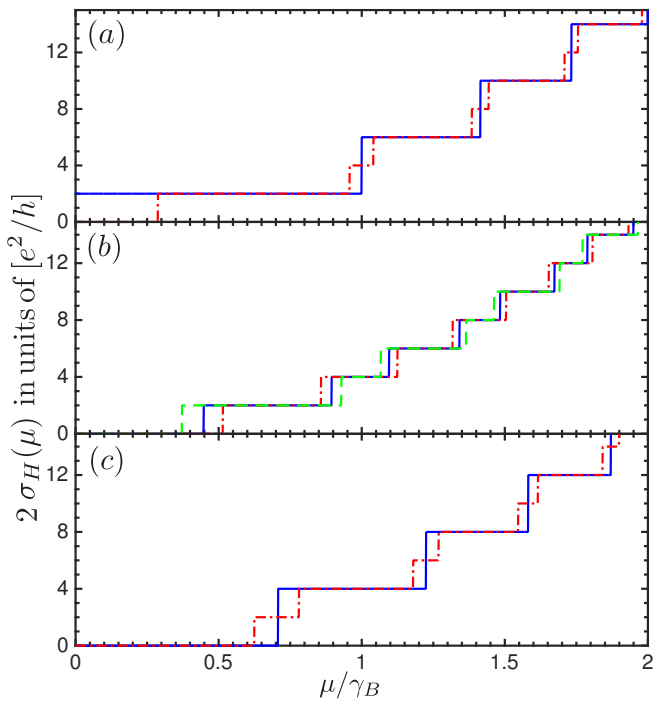}
\caption{(Color online) DC Hall conductivity curves plotting $2\sigma_H$ as a function of the chemical potential for various $\alpha$.  Here, we have included a factor of 2 to account for spin degeneracy.  Panel (a) shows $\alpha=0$ in solid blue, and $\alpha=0.3$ in red dot-dashed.  Panel (b) shows $\alpha=0.4,0.5,0.6$ in green dashed, solid blue and red dot-dashed, respectively.  Panel (c) shows $\alpha=1$ in solid blue and $\alpha=0.8$ in red dot-dashed.}
\label{fig:hall}
\end{figure}  
In this section, we derive the DC Hall conductivity of the $\alpha$-$T_3$ lattice from the grand potential via the magnetization.  To begin, we employ the Streda formula $e \frac{\partial M}{\partial\mu}=\sigma_H(\mu)$ to relate the Hall conductivity to the grand potential
\begin{align}\label{eq:partials}
	\sigma_{H}=e\frac{\partial M}{\partial \mu}=e\frac{\partial^2\Omega}{\partial\mu\partial B} = e\frac{\partial^2\Omega}{\partial B\partial\mu},
\end{align}
where we have used $M=-\frac{\partial\Omega}{\partial B}\Big|_{\mu}$ to obtain the magnetization from the grand potential.  Changing the order in which the partial derivatives are taken will allow us to take advantage of significant simplifications in later steps.

The relativistic grand potential can be found in Eq. (5.1) of Sharapov et al.~\cite{sharapov:2004} as
\begin{align}\label{eq:grand1}
\Omega(T,\mu)&=-T\int_{-\infty}^{\infty}N(\omega)\ln\left(2\cosh\left(\frac{\omega-\mu}{2T}\right)\right),
\end{align}
where $\mu$ is the chemical potential, the Boltzmann constant $k_B=1$, and $N(\omega)$ is the density of states in the magnetic field $B$.  Equation~\eqref{eq:grand1} can be written as 
\begin{align}\label{eq:grand2}
\Omega(T,\mu)=&-T\int_{-\infty}^{\infty}N(\omega)\ln\left(1+e^{\frac{\mu-\omega}{T}}\right)d\omega \\&+\frac{1}{2}\mu\int_{-\infty}^{\infty} N(\omega)d\omega -\frac{1}{2}\int_{-\infty}^{\infty}\omega N(\omega)d\omega. \nonumber
\end{align}
The first term of Eq.~\eqref{eq:grand2} is simply the grand thermodynamic potential in the non-relativistic case, while the last term is formally zero for the $\alpha$-$T_3$ model because it has particle-hole symmetry.  The second term will be retained for convenience but cannot contribute to the magnetization since it gives the total number of charge carriers in the system, and is independent of magnetic field.  In the limit of $T\rightarrow 0$, Eq.~\eqref{eq:grand2} gives
\begin{align}\label{eq:grand3}
\Omega(T=0,\mu)=&-\int_{-\infty}^{\infty}N(\omega)(\mu-\omega)\theta(\mu-\omega)d\omega\\
&+\frac{1}{2}\mu\int_{-\infty}^{\infty} N(\omega)d\omega, \nonumber
\end{align}
where $\theta(x)$ is the Heaviside function which is $0$ for $x<0$ and $1$ for $x>0$.  For $\mu>0$ the grand potential becomes
\begin{align}\label{eq:grand4}
\Omega(T=0,\mu)&=\int_{-\infty}^{0^-}N(\omega)(\omega-\mu)d\omega\\
&+\int_{0^+}^{\mu} N(\omega)(\omega-\mu)d\omega +\mu\int_{-\infty}^{0^-} N(\omega)d\omega, \nonumber
\end{align}
where particle hole symmetry $N(\omega)=N(-\omega)$ was used in the last term which cancels part of the first term to get 
\begin{align}\label{eq:grand5}
\Omega(T=0,\mu)=&\int_{-\infty}^{0^-}N(\omega)\omega d\omega\\
&+\int_{0^+}^{\mu} N(\omega)(\omega-\mu)d\omega \nonumber.
\end{align}
The first term in ~\eqref{eq:grand5} does not depend on the chemical potential but of course does depend on magnetic field through $N(\omega)$.  In this work we will be interested in the derivative of the magnetization with respect to the chemical potential taken at constant B (see Eq.~\eqref{eq:partials}).  Thus we can replace the $\Omega$ in this equation by only the second term in ~\eqref{eq:grand5}.  So we proceed with 
\begin{align}\label{eq:grand6}
\tilde{\Omega}(T=0,\mu)=&\int_{0^+}^{\mu} N(\omega)(\omega-\mu)d\omega.
\end{align}
and need to specify the density of states of the charge carriers under a magnetic field $ B $, which can be found in Eq.\eqref{eq:density1} of the previous section.  We ignore the flat band, as its energy does not depend on $B$ and hence it cannot make a contribution to the magnetization. Returning to Eq.~\eqref{eq:grand6} the derivative of $\tilde{\Omega}$ with respect to $\mu$ at constant $B$ gives
\begin{align}\label{eq:hall1}
	\frac{\partial\tilde{\Omega}(T=0,\mu)}{\partial\mu}\Bigg|_B=-\int_{0^+}^{\mu}N(\omega)d\omega.
\end{align}	
On substituting Eq.~\eqref{eq:density1} into the above, we find that $\frac{\partial\tilde{\Omega}_{\xi}}{\partial\mu}$ is $eB/h$ times an integer which depends on the chemical potential.  The value of the integer stays constant between LLs and increments by 1 each time $\mu$ crosses a new LL.  Thus, away from one of these jumps, taking the derivative with $B$ is equivalent to dropping the B in the expression for Eq.~\eqref{eq:hall1}.  The plateaux in the magnetization are related to the Hall plateaux by the Streda formula giving
\begin{align}\label{eq:hall2}
	\sigma_{H,\xi} & = \frac{e^2}{h}\left(\floor*{\frac{\mu^2}{\gamma_B^2}+\frac{1}{2}+\frac{\phi_{\xi}^B}{2\pi}}\right),
\end{align}	
which is one of our important results.  Here $\floor{x}$ is the integer part of $x$, limited to positive integers, as it counts levels crossed by $\mu$.  Eq.~\eqref{eq:hall2} never reduces to the formula for graphene given by Gusynin and Sharapov~\cite{gusynin:2005_1} (their Eq.~(2)).  To make the connection we need to include both valleys.  Combined we get 
\begin{align}\label{eq:hall4}
\sigma_{H}&=\sum_{\xi=\pm}\sigma_{H,\xi}=\frac{e^2}{h}\Big(\floor*{\frac{\mu^2c}{2v_F^2e\hbar B}+\frac{1}{2}+\frac{\phi^B}{2\pi}}\nonumber\\  
&+ \floor*{\frac{\mu^2c}{2v_F^2e\hbar B}+\frac{1}{2}-\frac{\phi^B}{2\pi}} \Big), 
\end{align}
where we have used a Berry phase $\phi^B$ which does not include the valley factor $\xi$ (ie, $\phi_{\xi}^B=\xi\phi^B$).  Representative cases for the Hall conductivity can be found in Fig.~\ref{fig:hall}.

Two limits are of particular interest: $\alpha=0$ and $\alpha=1$.  In the first case the Berry phase is $\phi^B=\pi$ while in the second case it is $0$.  For $\phi^B=\pi$ we get
\begin{align}\label{eq:hall5}
\sigma_{H}&=\frac{e^2}{h}\left( 1+2\floor*{\frac{\mu^2 c}{2v_F^2e\hbar B}}\right), 
\end{align}
which is Eq.~(7) of Gusynin and Shaparov~\cite{gusynin:2005_1} for graphene when a degeneracy factor of two for spin is accounted for.
Thus we obtain a relativistic series for the Hall quantization of $2,6,10,14,..$ in units of $e^2/h$, with a factor of two for spin degeneracy added for more convenient comparison to graphene.  This factor is not part of Eq.~\eqref{eq:hall4} and \eqref{eq:hall5} but is included in the top part of Fig.~\ref{fig:hall} (blue curve). 

For the other limiting case of zero Berry phase
\begin{align}\label{eq:hall6}
	\sigma_H&=\frac{e^2}{h}\left(2\floor*{\frac{\mu^2 c}{2v_F^2e\hbar B} +\frac{1}{2} } \right),
\end{align}
which is the quantization rule for the non-relativistic (classical) case.  The Hall series is now $0,4,8,12,...$ again in units of $e^2/h$ and with an extra factor of two added for spin.  This is shown in the bottom panel of Fig.~\ref{fig:hall} by the solid blue curve.

In general, for the relativistic case, the Hall plateaux are given by $\sigma_{xy}=2(2l+1)e^2/h$ with $l=0,1,2,...$ rather than the usual, non-relativistic QHE with $\sigma_{xy}=4le^2/h $.  This difference in the quantization of the Hall plateaux is traced~\cite{gusynin:2005_1} to the fact that in Dirac (relativistic) theory the degeneracy of the lowest energy Landau level (LL) is smaller than that of higher levels by a factor of two.  More specifically this arises because the lowest level is at zero energy and is consequently shared by the conduction and valence band.

For any intermediate value of $\alpha$ between $0$ and $1$ Eq.~\eqref{eq:hall4} applies, and provides the evolution as a function of $\alpha$ (or equivalently the Berry phase) of the Hall quantization from the relativistic to the non-relativistic series.  We illustrate this progression in Fig.~\ref{fig:hall}.  In the top panel, we show the case of finite, but small $\alpha$ ($\alpha=0.3$, red dot-dashed curve) superimposed over that of graphene ($\alpha=0$).  The effect of introducing the coupling $\alpha$ is quite dramatic.  From the form of the relativistic QHE of graphene, we see the emergence of a second set of Hall plateaux, resulting in quantization $0,2,4,6,...$, or half of the non-relativistic case.  

The spacing of the Hall plateaux is also unusual for small $\alpha$.  For $\alpha=0$ the first step (with height 2 in units of $e^2/h$) is quite long because a second LL is crossed only when the chemical potential equals $\gamma_B$.  In contrast to the $\alpha=0$ case, for small but finite $\alpha$, the lowest LL is no longer at zero energy.  It is rather at a finite $\mu^{\prime}=\gamma_B\sqrt{1/2-\phi^B/2\pi}$ (see Fig.~\ref{fig:berry}) so that $\sigma_H$ is zero at $\mu=0$ and does not jump to 2 until $\mu=\mu^{\prime}$.  Similarly, all higher energy LL's, which were two-fold degenerate for $\alpha=0$, become one fold degenerate for finite $\alpha$.  Thus they create a new series of steps with half the rise they had in the $\alpha=0$ case, but with very short runs.  In an experiment these new steps could be washed out by broadening due to scattering processes or temperature.  Note also that the run size of the various steps decreases with increasing $\mu$ so only the first few are likely to be seen.

In the general case of $\alpha$ finite and well away from the $0$ and $1$ limits, the onsets of new plateaux associated with either of the two valleys is more equally spaced in energy (see middle frame of Fig.~\ref{fig:hall}).  This eliminates the very short runs associated with the staircase at the top and bottom frames.  The three values of $\alpha=0.4$ (red dot-dashed), 0.5 (solid blue) and 0.6 (green dashed) all show similar run sizes given by $\gamma_B\sqrt{n-1/2-\phi_{\xi}^B/2\pi}$.  The final step of the evolution from relativistic to its non-relativistic counterpart is illustrated in the bottom frame of Fig.~\ref{fig:hall}, where we show the $\alpha=1$ case (blue) with quantization of $0,4,8,12,...$ and compare with $\alpha=0.9$ case (red dashed).  We see a second set of steps, the quantization now $0,2,4,6,...$ as in the general case (frame (b)) but the runs associated with the various steps alternate between very short and long in comparison.  The short steps disappear as $\alpha$ becomes 1.
  
Another related issue  associated with the magnetization which has been extensively studied is that of the connection between the Berry phase and the phase offset in the quantum oscillations associated with the Shubnikov-de Hass (SdH) and de Hass-van-Alphen (dHvA) effects.  In the non-relativistic case ~\cite{igor:2004}, with Berry phase of zero, the offset is 1/2 in unit of $2\pi$ while it is zero in the relativistic case even when the spectrum becomes gapped and the Berry phase is reduced from $\pi$~\cite{novoselov:2005,zhang:2005,sharapov:2004,gusynin:2005_2,fuchs:2010}.  A semi-classical discussion of phase offset and its relationship to the Maslov index and topological part of the Berry phase was given by J.N. Fuchs et al ~\cite{fuchs:2010} and further elaborated on in the case when the spectrum becomes gapped in references~\cite{wright:2013,tabert:2015_2}.  Here, the $\alpha$-$T_3$ model provides an example of a system that is different from the above.  It corresponds to a case where a non-relativistic Hall quantization is found, even though the phase offset in the oscillations is zero; a result previously found for the $S=1$ ($\alpha=1$) case~\cite{lan:2011,raoux:2014}.

\section{Dynamical Optical Conductivity}\label{sec:optics}
\begin{figure}[tbh]
\centering
 \includegraphics[width=\columnwidth]{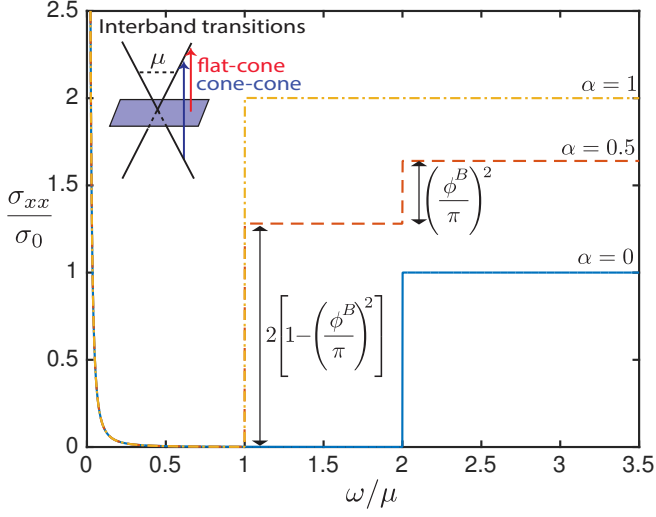}
\caption{(Color online) Optical conductivity curves for representative values of $\alpha$, including the case of graphene ($\alpha=0$) in solid blue and the $T_3$ lattice ($\alpha=1$) in yellow dot-dashed.  The Berry phase dependence of the conductivity is highlighted for the intermediate value of $\alpha=0.5$ shown in red-dashed.  For all three curves the Drude, centered at $\omega=0$ is broadened by $\gamma=0.001\mu$.  The inset is a schematic diagram depicting the set of possible interband transitions.}
\label{fig:optics}
\end{figure}
In this section, we calculate the dynamical optical conductivity of the $\alpha$-$T_3$ model at $B=0$ and highlight the role of the underlying Berry's phase in this quantity.  The optics associated with the two limiting cases of graphene and the dice lattice can be found in Ref.~\cite{dora:2011}. 

The absorptive part of the dynamical optical conductivity, $\sigma_{xx}(\omega)$, follows from the Kubo formula and is given in the pure limit as 
\begin{align}
	\sigma_{xx}(\omega)&= \sum_{s,s^{\prime}=0,\pm1}\sum_{\pmb{k},\pmb{k}^{\prime}}\frac{\theta(\mu-\varepsilon_{\pmb{k}^{\prime},s^{\prime}}^{\prime})-\theta(\mu-\varepsilon_{\pmb{k},s})}
	{\varepsilon_{\pmb{k},s}-\varepsilon_{\pmb{k}^{\prime},s^{\prime}}}\times\\
	&\pi\delta(\omega+\varepsilon_{\pmb{k},s}-\varepsilon^{\prime}_{\pmb{k}^{\prime},s^{\prime}})|\bra{\Psi_s}j_x\ket{\Psi_{s^{\prime}}^{\prime}}|^2\nonumber,
\end{align}	 
where the wavefunctions $\Psi_s$ are given in Section II by Eq.~\eqref{eq:wf1} and Eq.~\eqref{eq:wf2}.  The associated energies are $\varepsilon_{\pmb{k},0}=0$ for the flat band, and $\varepsilon_{\pmb{k},s}=s\hbar v_F k$ for the valence and conduction band.  The Heaviside step function $\theta(x)$ replaces the Fermi function for zero temperature.  The difference of the two $\theta$ functions ensures that only transitions that cross the chemical potential are permitted.  The current operator $j_x$ is defined as $j_x=-\frac{e}{\hbar} \frac{\partial\hat{H}}{\partial k_x}$, resulting in $j_x=-ev_FS_x$ for the case of interest.  The matrix $S_x$ has $\alpha$ dependence (note $\alpha=\tan(\varphi)$) and is
\begin{align}
	S_x&= \left(\begin{array}{c c c} 0 & \cosp & 0 \\
\cosp & 0 & \sinp \\ 
0 & \sinp & 0 \\
\end{array}\right).
\end{align}	
For simplicity, we will take $\mu$ $>$ $0$.  The required matrix elements are given as 
\begin{align}
&|\bra{\pm}j_x\ket{\pm}|^2=e^2v_F^2\cos^2\theta_{\pmb{k}}\nonumber\\	
&|\bra{\pm}j_x\ket{\mp}|^2=e^2v_F^2\sin^2\theta_{\pmb{k}}\costps\\
&|\bra{0}j_x\ket{\pm}|^2=|\bra{\pm}j_x\ket{0}|^2=\frac{e^2v_F^2}{2}\sin^2\theta_{\pmb{k}}\sintps\nonumber
\end{align}  
where we have used the shorthand $\bra{s}$ to denote $\bra{\Psi_s}$ with $s=\pm1$ for the conduction and valence band, respectively, and $s=0$ for the flat band.  

The first matrix element is responsible for the intraband contribution (Drude), while the latter two contribute to interband transitions.  Thus, the flat band does not contribute to the Drude weight since it has uniformly zero group velocity, but does contribute to the interband transitions.  We note that the second matrix element above is proportional to the square of the Berry phase of the cones, while the third is proportional to one minus the square of the same Berry phase.  The interband conductivity is found to be 
\begin{align}\label{eq:optics}
	\sigma_{xx}^{inter}(\omega)&=\sigma_0\left(\frac{\phi^B}{\pi}\right)^2\theta(\omega-2\mu)\nonumber\\
	&+2\sigma_0\left(1-\left(\frac{\phi^B}{\pi}\right)^2\right) \theta(\omega-\mu),
\end{align}	
with $\sigma_0=e^2/4\hbar$.  

There is also an intraband contribution which does not depend on the Berry phase and is 
\begin{align}\label{eq:sigma_intra}
	\sigma_{xx}^{intra}(\omega)&=4\sigma_0\mu\delta(\omega).
\end{align}	
For finite residual scattering, the $\delta$-function above broadens to a Drude $\delta(\omega)=\frac{1}{\pi}\frac{\gamma}{\omega^2+\gamma^2}$,	  
with scattering rate $\gamma$.  We note that there is a sum rule for conservation of optical spectral weight as the chemical potential $\mu$ is changed, where the weight lost in the interband background is transferred to the Drude centered about $\omega=0$.  
We write the dynamical conductivity as
\begin{align}\label{eq:full_optics}
	\sigma_{xx}(\omega)=&\sigma_0\costps\left( 4\mu\delta(\omega)+\theta(\omega-2\mu) \right)\nonumber\\
	+ &\sigma_0\sintps\left( 2[2\mu\delta(\omega)+\theta(\omega-\mu)] \right)
\end{align}
where we identify $(\phi^B/\pi)^2=\cos^2(2\varphi)$ as the square of the Berry phase, and $1-(\phi^B/\pi)^2=\sin^2(2\varphi)$ as one minus the square of the Berry phase.  This form of the conductivity emphasizes the connection with the expected conductivity of the two limiting cases.  The first line contains the graphene conductivity, multiplied by a Berry phase dependent factor, $\costps$.  This factor becomes $1$ for graphene.  The second line contains the conductivity associated with the dice lattice, multiplied by a different Berry phase dependent factor, $\sintps$.  Similarly, this factor becomes $1$ for the dice lattice.  The complete result is a superposition of these two limiting cases, where the dice result drops out for the graphene case, and vice versa.

The dynamical optical conductivity curves for a number of representative $\alpha$ values are shown in Fig.~\ref{fig:optics}.  An inset schematic is included to detail the interband transitions that are possible for positive chemical potential $\mu$, where arrows depict transitions from the lower to the upper Dirac cone and from the flat band to the upper Dirac cone.    

For the limiting case of $\alpha=0$, the sites of the HCL are decoupled from the central atom, leading to an inert flat band.  Thus, there are no flat band to cone transitions, and we observe no conductivity below twice the chemical potential.  Beyond that point, for $\omega>2\mu$, cone-to-cone transitions become possible and we find a constant conductivity of $\sigma_0$ (see the solid blue curve in Fig.~\ref{fig:optics}).  For $\alpha=0$ the first line of Eq.~\eqref{eq:full_optics} gives us the expected conductivity of graphene, and the second line drops out.  

For the other limiting case of $\alpha=1$, the interband conductivity is due entirely to flat band to cone transitions.  Here, we find a constant conductivity of $2\sigma_0$ for $\omega$ greater than the chemical potential $\mu$.  Below the chemical potential, the interband conductivity once again remains zero  (see the dot-dashed yellow curve in Fig.~\ref{fig:optics}).  In this case, the second line of Eq.~\eqref{eq:full_optics} gives us the conductivity of the dice lattice, and the first line drops out.  

In the intermediate regime, for $0<\alpha<1$, we see a superposition the two limiting cases described above (see the dashed red curve in Fig.~\ref{fig:optics}).  As before, below the chemical potential ($\omega<\mu$) there is no interband conductivity.  For $\mu<\omega<2\mu$ the conductivity is due entirely to flat band to cone transitions, and is given by $2[1-(\phi^B/\pi)^2] \sigma_0$.  For $\omega>2\mu$, cone to cone transitions become possible, and we see a step up of $(\phi^B/\pi)^2\sigma_0$ in the conductivity.  Thus, the intermediate regime is a mixing of the expected graphene and dice results (see Ref.~\cite{dora:2011} for examples of both), with a Berry phase dependent weight.  

It is important to note that in the two limiting cases, the Berry phase dependence of the conductivity is no longer explicitly obvious without prior knowledge of the formula we derived for the full range of $\alpha$.   
    
The dynamical nature of the optical conductivity allows it to access information about the underlying geometric structure of the lattice even without the presence of a magnetic field.  Thus, the dynamical optical conductivity is a $q\simeq 0$, finite $\omega$ quantity that knows about the Berry phase of the lattice.  

\section{Angular Scattering Probability}\label{sec:scattering}
\begin{figure}[bth]
\centering
 \includegraphics[width=0.8\columnwidth]{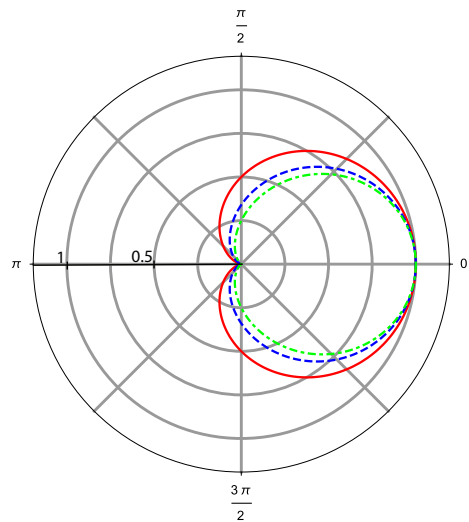}
\caption{(Color online) Angular scattering probability for zero magnetic field as calculated from Eq.~\eqref{eq:scattering1}.  Curves are shown for representative $\alpha$ values of $\alpha=0$, $\alpha=0.5$, $\alpha=1$ in solid red, dashed blue, and dot-dashed green, respectively.}
\label{fig:polar}
\end{figure}

As another example of a $B=0$ quantity that knows about the Berry phase, we calculate the angular scattering probability for the $\alpha$-$T_3$ model. This is given by the square of the matrix element overlap in the conduction band between the initial state $\ket{\Psi(\pmb{k})}$ and final state $\ket{\Psi(\pmb{k}^{\prime})}$ with $|\pmb{k}|=|\pmb{k}^{\prime}|$.  The wavefunction for the conduction band is
\begin{equation}
\ket{\Psi(\pmb{k})} =\frac{1}{\sqrt{2}}\left(\begin{array}{c } \cosp\, e^{i\theta_{\pmb{k}}} \\
 1 \\ 
\sinp\, e^{-i\theta_{\pmb{k}}}\\
\end{array} \right).\\
\end{equation}
Here, $\theta_{\pmb{k}}$ is the angle associated with momentum $\pmb{k}$ such that $f_{\pmb{k}}=|f_{\pmb{k}}|e^{i\theta_{\pmb{k}}}$.  Thus, the scattering probability works out to be 
\begin{align}\label{eq:scattering1}
	|\branobar{\Psi(\pmb{k})}\ket{\Psi(\pmb{k}^{\prime}}|^2&=\left(\frac{1+\cos{\theta_{\pmb{k}^{\prime}}}}{2}\right)^2+\left(\frac{\phi^B}{2\pi}\right)^2\sin^2\theta_{\pmb{k}^{\prime}}
\end{align}	
where we take $\pmb{k}$ is along the $x$-axis and $\theta_{\pmb{k}^{\prime}}$ is the scattering angle with respect to this initial state.  

In graphene, the chiral nature of the quasiparticles results in a complete suppression of backscattering~\cite{ando:1998,mceuen:1999}, as a consequence of pseudospin conservation.  For the $\alpha$-$T_3$ model we note that the Berry phase remains in Eq.~\eqref{eq:scattering1} as shown in Fig.~\ref{fig:polar}, and modulates the scattering probability by the square of the Berry phase for angles of $\theta_{\pmb{k}}\neq \pi m$, for integer $m$.  As in graphene, backscattering is absent for all $\alpha$.  Similarly, forward scattering remains unchanged with probability of 1, independent of $\alpha$.  However, for all other angles, angular scattering is suppressed with decreasing Berry phase, though still finite even for a Berry phase of zero (dice lattice).  

The matrix element considered in Eq.(32) can be utilized to calculate both quasiparticle and transport scattering rates, $\tau^{-1}_{qp}$ and $\tau^{-1}_{tr}$, respectively, in the case of a scattering potential which is independent of momentum transfer.  The latter is weighted by an additional factor of $(1-\cos\theta_{\mathbf{k^{\prime}}})$, which encodes the information that forward scattering does not deplete the current, while backward scattering has maximal effect.  Upon averaging Eq.(32) over angles $\theta_ {\mathbf{k^{\prime}}}$ with and without the inclusion of the transport factor $(1-\cos\theta_{\mathbf{k^{\prime}}})$, we can determine the ratio of the quasiparticle to transport scattering rates.  This ratio is given by
\begin{align}
	\frac{\tau_{tr}}{\tau_{qp}}&=\frac{3+\left(\frac{\phi^B}{\pi}\right)^2}{1+\left(\frac{\phi^B}{\pi}\right)^2}\nonumber .
\end{align}	
For $\phi^B=\pi$, this is equal to $2$, the standard result found for graphene.  Here we find that it increases to 3 in the $\alpha$-$T_3$ model for $\phi^B=0$.

The angular scattering probability is a zero $\omega$, finite $\pmb{q}=\pmb{k}^{\prime}-\pmb{k}$ scattering process that is affected by the underlying Berry phase. 

\section{Conclusion}\label{sec:conclusion}
In this paper we utilized the variable Berry phase of the $\alpha$-$T_3$ model to highlight the effect of this phase on a number of physical observables.  

In the density of states, an inherently static quantity, we noted that the Berry phase is only visible in the presence of a motion-inducing magnetic field.  In the limit of zero field, the Berry phase drops out of the density of states, resulting in the same (dimensionless) density of states for the conical bands, independent of the parameter $\alpha$.    

We took advantage of the variable Berry phase of the $\alpha$-$T_3$ model to examine the evolution of the Hall quantization from a relativistic to a non-relativistic sequence.  As we varied $\alpha$ from $0$ to $1$, we noted the emergence, followed by gradual increase in lengths, of plateaux (plotted as a function of the chemical potential) that would form the non-relativistic quantization rule.  This was accompanied by the simultaneous gradual decrease, and finally, disappearance of the plateaux that had formed the original, relativistic quantization rule.  

Dynamical quantities are effected by the underlying Berry phase, even in the absence of a motion inducing magnetic field.  We presented examples of two $B=0$ quantities, one with zero $q$ and finite $\omega$ (optical conductivity), and one with zero $\omega$ and finite $q$ (angular scattering probability), that demonstrate knowledge of the Berry phase.  We showed that the absorptive part of the longitudinal conductivity for this model can be separated into two Berry phase dependent functions.  Both of these functions play a role in the intermediate regime ($0<\alpha<1$), but for the two limiting cases (graphene and dice lattice), one of the two functions drops out, making the Berry phase dependence no longer explicitly obvious without prior knowledge of the intermediate regime.  The angular scattering probability was also shown to be a function that explicitly depends on the underlying Berry phase. 

The result that one can see changes in these $B=0$ and finite $B$ quantities due to the variation of the Berry phase is a manifestation of the fact that the Berry phase is a geometrical phase.  It is clear that in a magnetic field the electrons will be forced to circulate in a semiclassical sense and hence will know about the Berry phase.  But, even the $B=0$ properties discussed here will be sensitive to the phase as the electromagnetic probe of light in optics drives the electrons to motion, and angular scattering probability also implies motion for electrons scattering from $\mathbf{k}$ to $\mathbf{k^{\prime}}$.

\section{Acknowledgements}
We acknowledge G. Demand, J.D. Malcolm and C.J. Tabert for useful discussions.  This work has been supported by the Natural Science and Engineering Research Council (NSERC) of Canada, and in part, by the Canadian Institute for Advanced Research (CIFAR).


	
\end{document}